\documentclass[%
 aip,
 amsmath,amssymb,
 reprint,%
]{revtex4-1}

\usepackage{graphicx}
\usepackage{dcolumn}
\usepackage{bm}

\usepackage[utf8]{inputenc}
\usepackage[T1]{fontenc}
\usepackage{mathptmx}
\usepackage{etoolbox}
\usepackage{here}
\usepackage{float}

\makeatletter
\def\@email#1#2{%
 \endgroup
 \patchcmd{\titleblock@produce}
  {\frontmatter@RRAPformat}
  {\frontmatter@RRAPformat{\produce@RRAP{*#1\href{mailto:#2}{#2}}}\frontmatter@RRAPformat}
  {}{}
}%
\makeatother
\begin{document}

\preprint{AIP/123-QED}

\title[Passive Time-Varying Waveform-Selective Metasurfaces for Attainment of Magnetic Property Control]{Passive Time-Varying Waveform-Selective Metasurfaces for Attainment of Magnetic Property Control}
\author{Yuki Kunitomo}
\author{Kairi Takimoto}
\affiliation{ 
Department of Engineering, Nagoya Institute of Technology, Aichi 466-8555, Japan
}%

\author{Stefano Vellucci}%
\affiliation{
Department of Engineering, Niccol\`{o} Cusano University, Via Don Carlo Gnocchi 3, 00166, Rome, Italy
}%

\author{Alessio Monti}%
\affiliation{
Department of Industrial, Electronic and Mechanical Engineering, Roma Tre University, Via Vito Volterra 62, 00146, Rome
}%

\author{Mirko Barbuto}%
\affiliation{
Department of Engineering, Niccol\`{o} Cusano University, Via Don Carlo Gnocchi 3, 00166, Rome, Italy
}%

\author{Alessandro Toscano}%
\author{Filiberto Bilotti}%
\affiliation{
Department of Industrial, Electronic and Mechanical Engineering, Roma Tre University, Via Vito Volterra 62, 00146, Rome
}%

\author{Hiroki Wakatsuchi}%
 \email{wakatsuchi.hiroki@nitech.ac.jp.}
\affiliation{
Department of Engineering, Nagoya Institute of Technology, Aichi 466-8555, Japan
}%

\date{\today}

\begin{abstract}
We present circuit-loaded metasurfaces that behave differently in a passive manner even at the same frequency in accordance with the incoming waveform, specifically, its pulse width. Importantly, the time-varying waveform-selective metasurfaces reported thus far were mostly able to change their electric properties but not their magnetic properties; this severely limited the design range of their corresponding wave impedances and refractive indices and thus hindered the development of potential applications in antennas, sensors, imagers, signal processing, and wireless communications. In this study, passive time-varying waveform-selective metasurfaces were found to attain magnetic property control by introducing an additional circuit-loaded layer that generated an artificial magnetic dipole moment; this magnetic moment only occurred during the designed pulse duration in the time domain. Our proposed concept and structures were validated numerically and experimentally; thus, our results could be used to address electromagnetic and related issues sharing the same frequency component via the variation of the pulse width as an additional degree of freedom. 
\end{abstract}

\maketitle

\section{Introduction}
Electromagnetic phenomena are readily controlled by artificially engineered subwavelength structures referred to as metamaterials \cite{MTMbookEngheta,calozBook} and metasurfaces.\cite{EBGdevelopment, yu2014flat, chen2016review} For instance, metamaterials show exotic properties and characteristics, resulting in negative refraction,\cite{smithDNG1D, smithDNG2D1} diffraction limit breaking,\cite{pendryperfetLenses, lerosey2007focusing} antenna miniaturization,\cite{ziolProcIEEE} cloaking,\cite{pendryCloaking, schurig2006metamaterial, enghetaCloaking} ultra-thin absorbers,\cite{mtmAbsPRLpadilla, ultraThinAbs, My1stAbsPaper} and so on. Metasurfaces are a planar class of metamaterials and enable abrupt phase and magnitude changes of incoming waves, leading to wavefront control or beamforming over the simplified two-dimensional surfaces.\cite{yu2011light, pfeiffer2013metamaterial} Superior performance is attained by introducing nonlinearity into metamaterials and metasurfaces.\cite{lapine2014colloquium} Metasurfaces loaded with nonlinear circuit components show power-dependent states; thus, they effectively protect the sensitive electronic devices from high-power electromagnetic noise, and maintain small-signal telecommunications at the same frequency.\cite{aplNonlinearMetasurface, katko2011rf, kim2016switchable, li2017high} These electromagnetic property changes provide a higher degree of freedom to design electromagnetic phenomena, which has been widely and intensively explored in recent years as time-varying metamaterials and metasurfaces from microwave to optical frequencies.\cite{won2021new} A simple approach for achieving time-varying responses includes introducing nonlinear media/circuits with a biasing system.\cite{THzActiveMTMpadilla, shrekenhamer2013liquid, zhang2018space} Time-varying metasurfaces provide interesting features such as non-reciprocity\cite{caloz2018electromagnetic, wakatsuchi2019numerical, li2020time}, artificial Doppler effect,\cite{ramaccia2019phase} and pulse shaping.\cite{chamanara2019simultaneous} However, most conventional time-varying metasurfaces require an external energy supply, which limits the applicable scenarios. In contrast, circuit-based metasurfaces were reported to show variations in their responses at the same frequency in accordance with the incoming waveform, specifically, the pulse width.\cite{wakatsuchi2013waveform, eleftheriades2014electronics, wakatsuchi2019waveform, cheng2023waveform} These waveform-selective metasurfaces have an advantage of a biasing-circuit-free design, which is already exploited in noise suppression,\cite{wakatsuchi2019waveform} antenna design,\cite{vellucci2019waveform, barbuto2020waveguide, barbuto2021metasurfaces, ushikoshi2023pulse}, sensing \cite{ushikoshi2023pulse}, signal processing \cite{f2020temporal}, and wireless communications.\cite{wakatsuchi2015waveformSciRep, takimoto2023perfect} In particular, their passive time-varying response can be utilized as intelligent reflecting surfaces (IRSs) or reconfigurable intelligent surfaces (RISs) \cite{basar2019wireless, tang2020wireless, ino2023noncoherent} since wireless communication systems no longer require biasing systems and precise symbol level synchronization with transmitters.\cite{fathnan2022method, fathnan2023unsynchronized} Moreover, waveform-selective metasurfaces operating at multiple frequencies \cite{takeshita2023dual, qian2024design} exhibit a higher degree of freedom to distinguish different signals with the same frequency resources in accordance with the incoming frequency sequence.\cite{takeshita2024frequency} However, conventional waveform-selective metasurfaces only showed time-varying response for control of their electric properties but not their magnetic properties. Therefore, these metasurfaces produced a limited range of electromagnetic properties; these properties were related to the wave impedance and refractive index and thus limited the degree of freedom to control electromagnetic phenomena. For this reason, in this study, passive time-varying waveform-selective metasurfaces that could change their magnetic properties in accordance with the pulse width of the incident wave were developed. The integration of the magnetic property control with waveform-selective metasurfaces allows 100 \% control of the wave impedance and refractive index and thus expands their potential application scenarios, such as switching between opaque and transparent windows, thereby leading to significant performance enhancement in antennas, sensors, imagers, and wireless communications. Note that in this study we abbreviate 50-ns pulsed wave and continuous wave as "PW" and "CW", respectively, each of which is used to evaluate the initial state and the steady state of our metasurfaces. 

\begin{figure*}[tb!]
\includegraphics[width=\linewidth]{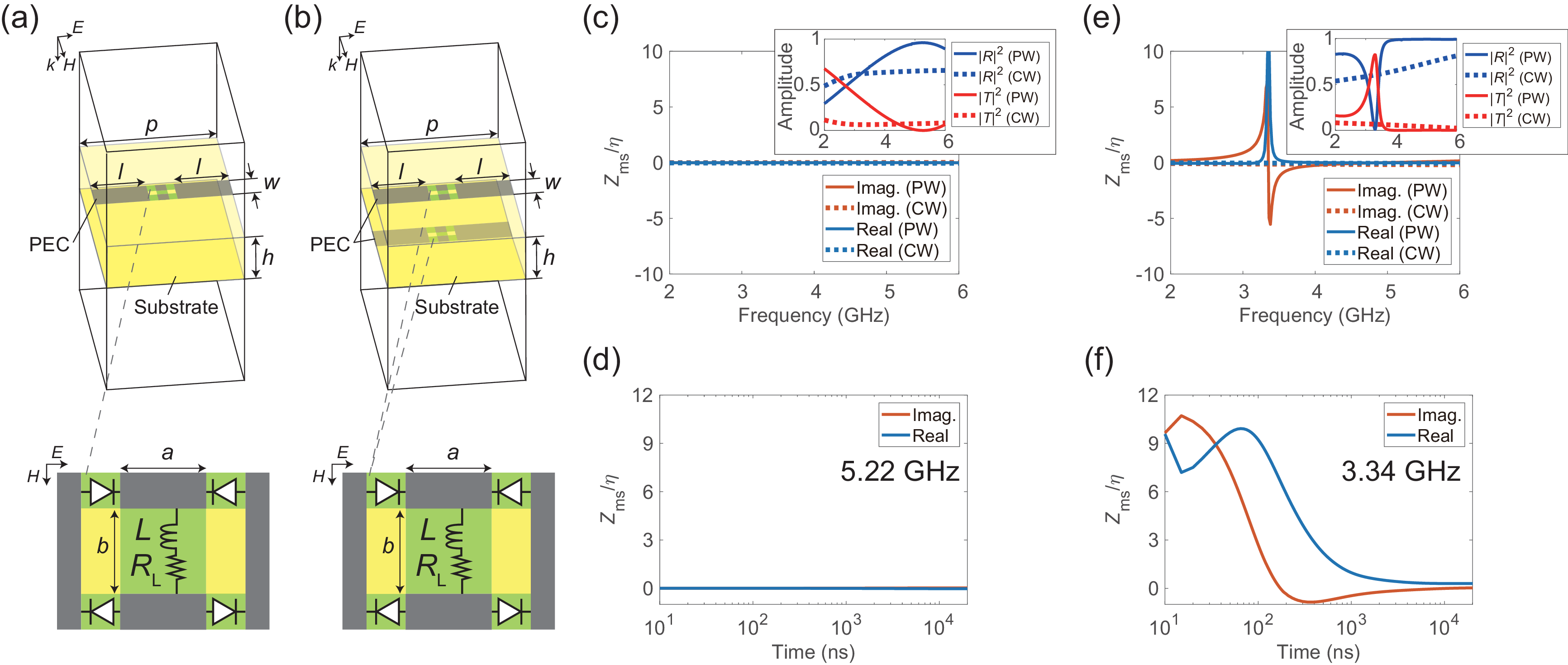}
\caption{\label{fig1}Conventional and proposed models and their simulation results. (a) Conventional one-layer cut-wire-based waveform-selective metasurface and loaded circuit configuration containing an inductor connected to a resistor in series. Design parameters were as follows: $p=$ 17.00 mm, $l=$ 5.00 mm, $h=$ 0.64 mm, $w=$ 3.50 mm, $a=$ 3.00 mm, $b=$ 1.50 mm, $L=$ 1 mH, and $R_L=$ 24 $\Omega$. (b) Proposed two-layer cut-wire-based waveform-selective metasurface and loaded circuit configuration. The model adopted the same design parameters as those used for the conventional model except for $l, h,$ and $w$ that were changed to 6.00, 3.84, 9.00 mm, respectively, to enhance the magnetic response (see Fig.\ \ref{fig:figS0a} in Supplementary Material for the result of the original dimensions). (c) Simulated scattering parameters of the conventional one-layer model (inset) and its corresponding magnetic impedance $Z_{ms}$. (d) $Z_{ms}$ variation of the conventional one-layer model at 5.22 GHz in the time domain. (e) Simulated scattering parameters of the proposed two-layer model (inset) and its corresponding $Z_{ms}$. (f) $Z_{ms}$ variation of the proposed two-layer model at 3.34 GHz in the time domain. }
\end{figure*}

\section{Theory}

The fundamental design theory of waveform-selective metasurfaces has been reported in the literature \cite{wakatsuchi2013waveform, eleftheriades2014electronics, wakatsuchi2019waveform, cheng2023waveform} and is briefly understood as follows. In short, waveform-selective metasurfaces contain not only ordinary resonant subwavelength unit cells but also rectifying circuits that control the degree of the resonant intensity according to the transients well known in direct-current (DC) circuits. For instance, as a typical waveform-selective metasurface \cite{wakatsuchi2019waveform}, our metasurface unit cell was designed with a cut wire (12 $\times$ 9 mm$^2$) and a dielectric substrate (Rogers3010) as shown in Fig.\ \ref{fig1}a. Note that our metasurface design had small conducting pads between the cut wires to realistically deploy the circuit components (see the gap between cut wires in Fig.\ \ref{fig1}a). Without any circuit components, the metasurface strongly reflected the incoming wave at a designed resonant frequency.\cite{dolling2005cut, cwFiltering} This resonant mechanism was prevented by a diode bridge that rectified electric charges induced by the incident wave (following a sine function) to generate a fully rectified current (following a modulus of the same sine function). This waveform rectification converted the incoming frequency component to an infinite set of frequencies. However, most of the incident energy appeared at zero frequency as shown by the Fourier expansion of $|$sin$|$.\cite{wakatsuchi2019waveform} For this reason, the transient of the DC circuits could be made available within the metasurface and was associated with its resonant mechanism if an additional reactive component was included within the diode bridge in combination with a resistor. Specifically, we introduced an inductor of $L=$ 1 mH in series with a resistor of $R_L=$ 24 $\Omega$ as shown in the bottom of Fig.\ \ref{fig1}a. Under this circumstance, the reflecting resonant mechanism of the cut wire was still maintained if the incident waveform was sufficiently short since the inductor prevented incoming induced electric charges due to its strong electromotive force. However, this force was gradually reduced due to the zero frequency component of the induced electric charges. Therefore, when the incoming waveform continued as a sufficiently long pulse or a CW, the gap between the cut wires was short-circuited such that the intrinsic resonant mechanism of the metasurface was weakened, resulting in poor reflectance magnitude at the same frequency. 

However, this single-layer design permits only electric property control since this design does not generate an artificial magnetic dipole moment.\cite{fathnan2022method} To address this issue, we introduced another layer to deploy the same cut wire as shown in Fig.\ \ref{fig1}b. In this structure, the same circuit configuration as the one adopted in the conventional one-layer model was used for both the top and bottom layers. For this reason, without any additional circuit elements, the metasurface showed an artificial magnetic dipole moment.\cite{SoukoulisMTMs} This magnetic resonant mechanism, however, could be controlled by the above-mentioned transient well known in DC circuits. Therefore, the magnetic resonance, i.e., the magnetic property control, was maintained during PW illumination but was weakened in the steady state even at the same frequency. 

First, we simulated these two structures using a co-simulation method integrating an electromagnetic solver with a circuit simulator.\cite{wakatsuchi2019waveform, ushikoshi2023pulse, takeshita2024frequency} This simulation method was conducted by the ANSYS Electronics Desktop Simulator (2023 R2). This method was incapable of fully visualizing field distributions,\cite{wakatsuchi2015fieldVisualisation} while effectively simulating the transient response of the electromagnetic simulation models including the nonlinear circuit components in the time domain. Thus, the co-simulation method aided in the efficient determination of the suitable design parameters.

The conventional and proposed models were tested with two types of waveforms: PWs and CWs. In the former case, the incident power was fully integrated in the time domain as the incident energy and then compared with the reflected and transmitted energies to calculate the reflectance and transmittance. In the latter case, we used a harmonic balance approach to readily obtain a steady-state response in the time domain; here, the incident, reflected, and transmitted energies were similarly compared to provide the reflectance and transmittance. The reflectance and transmittance were then used to calculate magnetic impedance $Z_{ms}$; this represents the homogeneous impedance and is equivalent to the impedance of the original rigorous model through the following equation: \cite{holloway2005reflection, pfeiffer2013metamaterial, selvanayagam2013circuit, fathnan2022method} 
\begin{eqnarray}
Z_{ms}=\frac{2\eta(1-T+R)}{1+T-R}
\label{eq1},
\end{eqnarray}
where $\eta$, $T$, and $R$ represent the wave impedance of the vacuum, the transmission coefficient, and the reflection coefficient, respectively. Thus, transmittance and reflectance were obtained by $T^2$ and $R^2$, respectively, in our study (i.e., not $T$ and $R$). In addition, we also analyzed the transient responses of our simulation models. In this case, the incident, reflected, and transmitted powers were integrated over 20 ns in the time domain and updated every 5 ns to derive the transient reflectance and transmittance; these were used to calculate time-varying $Z_{ms}$ by Eq.\ (\ref{eq1}). 

\section{Results}

Under these circumstances, we calculated the scattering parameters and magnetic impedance of our models as shown in Fig.\ \ref{fig1}c to Fig.\ \ref{fig1}f. First, as shown in the inset of Fig.\ \ref{fig1}c, the conventional one-layer model resonated and strongly reflected the incident wave near 5.22 GHz. However, since this structure did not produce an artificial magnetic dipole moment, the normalized $Z_{ms}$ remained nearly unchanged for both PWs and CWs. This result is also clearly demonstrated in Fig.\ \ref{fig1}d; here, $Z_{ms}$ did not show variation in the time domain even at the resonant frequency of 5.22 GHz. In contrast, $Z_{ms}$ was varied in our proposed two-layer model that exhibited an additional resonant mechanism near 3.34 GHz, as shown in the inset of Fig.\ \ref{fig1}e. Note that in this structure we varied the cut-wire width and length as well as the substrate thickness to enhance the magnetic property control (see the caption of Fig.\ \ref{fig1} for detailed dimensions and Fig.\ \ref{fig:figS0a} of Supplementary Material for the response of the original structure). Under this circumstance, the reflectance magnitude at 3.34 GHz varied depending on the incident waveform unlike the result of the conventional one-layer model (cf.\ the inset of Fig.\ \ref{fig1}c). Additionally, the reflectance for the PW was suppressed at this frequency, while the transmittance was enhanced. These results signify the presence of the magnetic resonance, which is more clearly shown in the main panel of Fig.\ \ref{fig1}e; here, the imaginary part of $Z_{ms}$ largely changed for the PW but not for the CW at the same frequency. These results indicate that the proposed structure achieved the magnetic property control that appeared during PW illumination but eventually disappeared in the steady state. This trend is further clarified in Fig.\ \ref{fig1}f. In this figure, the real part of the magnetic impedance once decreased and increased between 10 and 100 ns presumably because the two circuit configurations deployed in the first and second layers did not simultaneously respond to the incident wave. Nevertheless, Fig.\ \ref{fig1}f shows that the imaginary part of the magnetic impedance at 3.34 GHz was approximately 11 times higher than the wave impedance of vacuum during the initial period of time, and the impedance gradually decreased to zero by increasing the incident pulse width. 

\begin{figure}[tb!]
\includegraphics[width=\linewidth]{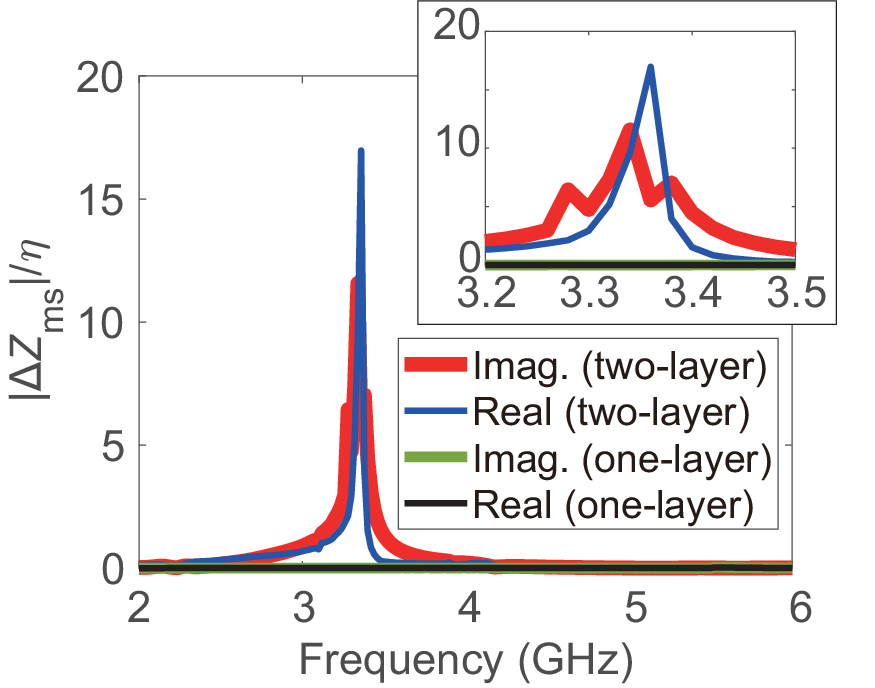}
\caption{\label{fig2}Magnetic impedance change in the proposed (two-layer) and the conventional (one-layer) waveform-selective metasurfaces. The inset represents the results near 3.34 GHz. }
\end{figure}

Next, we evaluated the degree of the magnetic property change of the proposed model with that of the conventional model in Fig. \ref{fig2}. Here, we assessed the change in the magnetic surface impedance, $\Delta Z_{ms}$ by subtracting the maximum value of $Z_{ms}$ by the minimum value of $Z_{ms}$ in the time domain at each frequency. As shown in Fig.\ \ref{fig2}, the conventional one-layer model varied the imaginary part of $Z_{ms}$ up to approximately 0.029 between 2 and 6 GHz. This change was significantly increased by a factor of 400 in the case of the proposed two-layer model (see near 3.34 GHz in the inset of Fig.\ \ref{fig2}). 

\begin{figure}[tb!]
\includegraphics[width=\linewidth]{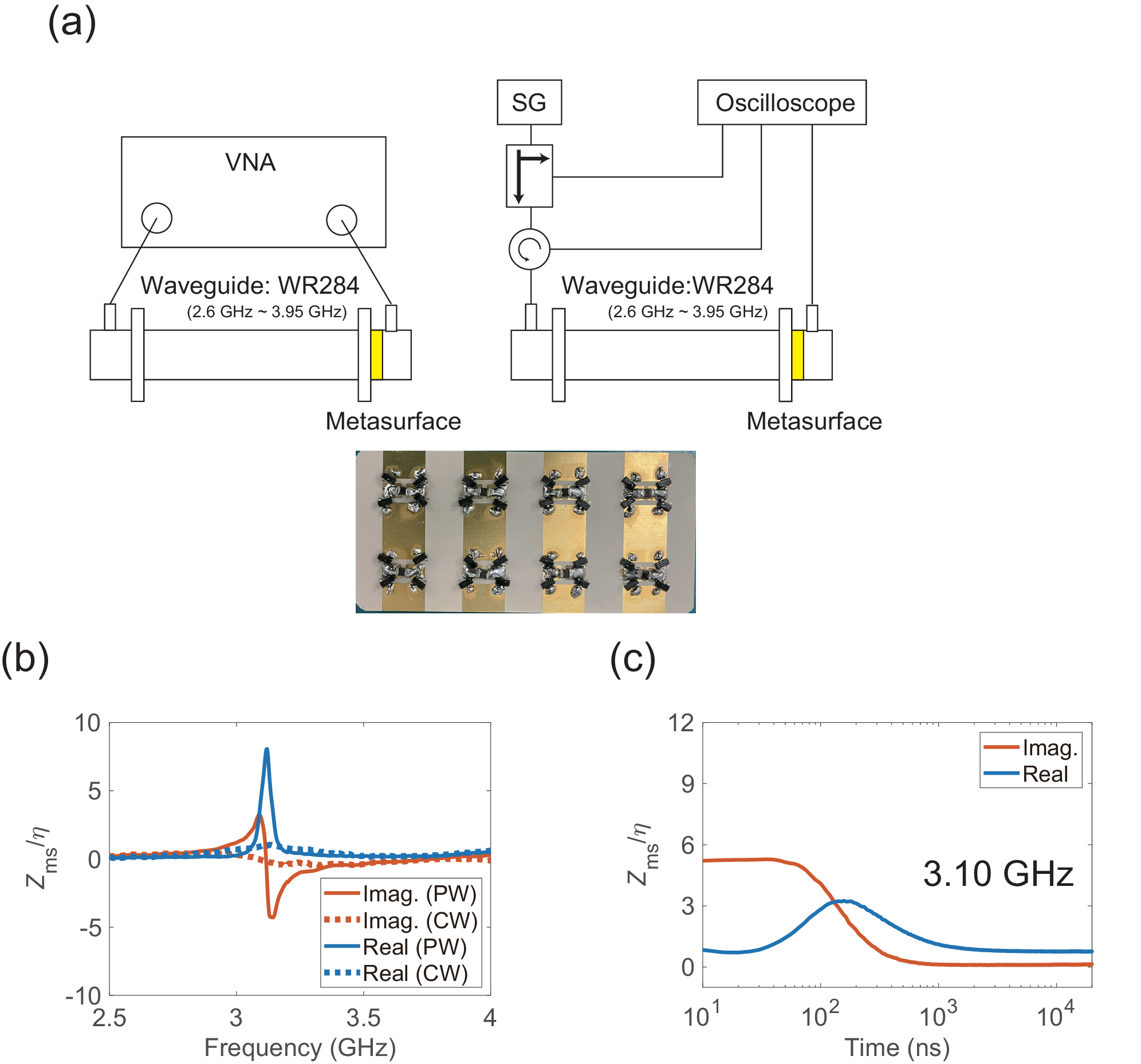}
\caption{\label{fig3}Experimental validation of the proposed structure. (a) Measurement setups and sample. (b, c) $Z_{ms}$ measured in (b) the frequency domain and (c) the time domain at 3.11 GHz. The input power was set to 13 and 15 dBm in the frequency- and the time-domain measurements, since the maximum power available for the frequency-domain measurement was limited to 13 dBm. }
\end{figure}

The proposed two-layer structure was then experimentally demonstrated, as shown in Fig. \ref{fig3}. In these measurements, we used the same design parameters as those adopted in the simulation model of Fig.\ \ref{fig1}b and fabricated a two-layer waveform-selective metasurface sample comprising of 2 $\times$ 4 unit cells as shown in Fig.\ \ref{fig3}a. The measured sample was then deployed inside a standardized rectangular wave (WR284) that was connected to either a vector network analyzer (VNA) (Keysight Technologies, N5249B) or a combination of a signal generator (SG) (Anritsu, MG3690C) and an oscilloscope (Teledyne Lecroy, 9404M) to obtain the frequency-domain or time-domain responses, respectively. Note that PWs and CWs were generated in the VNA by switching a pulse generation mode, while the SG simply increased the pulse width to observe the sufficiently long time-domain response of our metasurface. First, the frequency-domain $Z_{ms}$ was obtained as shown in Fig.\ \ref{fig3}b, where the imaginary part of $Z_{ms}$ substantially varied around 3.10 GHz. Thus, the measured result was consistent with the simulated result of Fig.\ \ref{fig1}e, although a minor frequency shift appeared due to some differences with the simulation setup; these differences could include additional parasitic parameters in the measurement sample, lateral boundary conditions, and the angle of the incident wave. Despite these differences, our measurement result supported that the proposed two-layer waveform-selective metasurface was capable of varying the magnetic impedance even at the same frequency in accordance with the incoming waveform. This characteristic is more clarified in Fig.\ \ref{fig3}c, where the imaginary part of $Z_{ms}$ changed approximately from 5.28 to 0.10 at 3.10 GHz in the time domain; this result also agreed with the simulation result of Fig.\ \ref{fig1}f. 

\begin{figure*}[tb!]
\includegraphics[width=\linewidth]{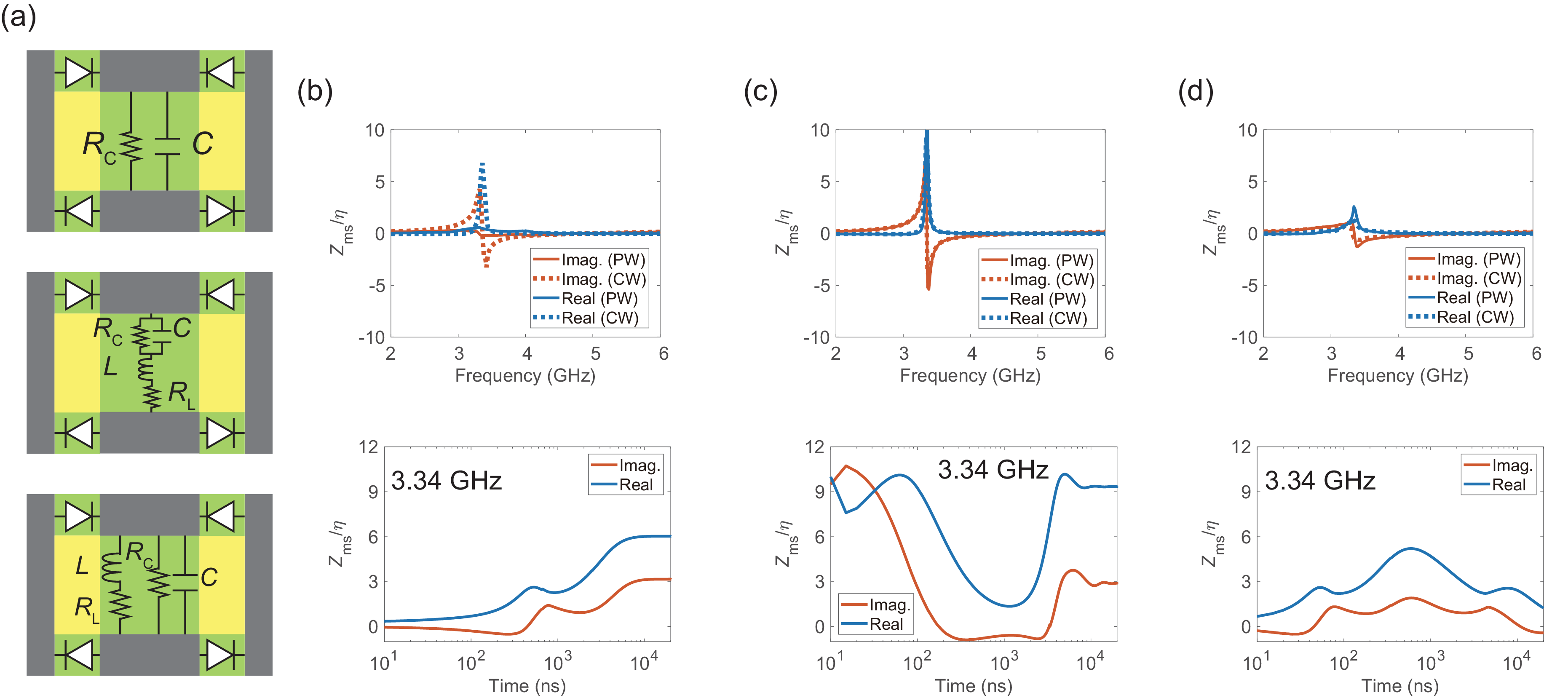}
\caption{\label{fig4}Additional simulation models to achieve various pulse width dependences. (a) Circuit configurations adopted for (top) capacitor-based, (middle) series-type, and (bottom) parallel-type waveform-selective metasurfaces. (b-d) Simulated $Z_{ms}$ results for (b) capacitor-based, (c) series-type, and (d) parallel-type waveform-selective metasurfaces. }
\end{figure*}

Although the results demonstrated above (e.g., Fig.\ \ref{fig3}c) showed that the imaginary part of $Z_{ms}$ gradually decreased, we also provided a design approach that arbitrarily increased the imaginary part of $Z_{ms}$ or produced more complicated $Z_{ms}$ by using the additional three circuit configurations shown in Fig.\ \ref{fig4}a.\cite{wakatsuchi2015waveformSciRep} First, the structure based on a capacitor of $C=$ 1 nF in parallel with a resistor of $R_C=$ 10 k$\Omega$ (top of Fig.\ \ref{fig4}a) allowed induced electric charges of a PW to enter the diode bridge. Therefore, the resonant mechanism to strongly reflect the incident wave was suppressed, as shown in Fig.\ \ref{fig4}b. However, since the capacitor was fully charged up in the steady state, a large imaginary part of $Z_{ms}$ appeared by increasing the incident pulse width as opposed to Fig.\ \ref{fig1}e and Fig.\ \ref{fig1}f. Second, the two types of the circuit configurations (i.e., the inductor- and the capacitor-based ones) were connected to each other in series within a diode bridge as a series-type waveform-selective metasurface (middle of Fig.\ \ref{fig4}a). By setting the circuit values to $L=$ 1 mH, $C=$ 1 nF, $R_L=$ 24 $\Omega$, and $R_C=$ 10 k$\Omega$, this structure only caused a decrease in its magnetic resonant mechanism for an intermediate time scale and maintained the magnetic resonance for short and long time scales; this result is shown in Fig.\ \ref{fig4}c. Third, the inductor- and the capacitor-based circuit configurations could also be combined in parallel as a parallel-type waveform-selective metasurface (bottom of Fig.\ \ref{fig4}a). In this case, the imaginary part of $Z_{ms}$ became large during an intermediate time slot by setting the circuit values to $L=$ 10 mH, $C=$ 0.1 nF, $R_L=$ 24 $\Omega$, and $R_C=$ 10 k$\Omega$, since the metasurface did not short-circuit the gap between the cut wire edges during this time frame. 

The proposed design approach for passive metasurfaces successfully caused the variation of the magnetic impedance at the same frequency in accordance with the incident pulse width. In particular, the magnetic impedance was more arbitrarily designed by properly selecting a circuit configuration within a diode bridge. Note that the circuit values were very important in characterizing the time-domain response. In Fig.\ \ref{fig:figS1a} of Supplementary Material, we provided additional results to show how the circuit values were related to $Z_{ms}$ in the time domain. The Supplementary Material also provided the measurement results of the capacitor-based waveform-selective metasurface (see Fig.\ \ref{fig:figS2a}). Importantly, the proposed design approach can be integrated with another metasurface element that varies an electric property in the time domain; this would enable access to the full range of the wave impedance and refractive index values. In this case, for instance, the metasurface response can be 100 \% changed from a fully reflecting state to a transmitting state; this can be utilized to develop applied microwave devices, such as antennas, sensors, and imagers and broaden their potential applications. 

\section{Conclusion}

In conclusion, we developed passive time-varying waveform-selective metasurfaces that were capable of varying their magnetic properties. The proposed structures had unit cells composed of cut wires connected by lumped circuit elements that converted the incoming frequency component to generate a large DC component; thus, these structures exhibited transient responses even at the same frequency in accordance with the incident pulse width. However, a one-layer cut-wire design did not greatly vary the magnetic impedance since the structure did not generate an artificial magnetic dipole moment, which was successfully produced by introducing another cut-wire layer. Based on our results, the imaginary part of the magnetic impedance of the proposed structure was approximately 11 and 5 times larger than the wave impedance of a vacuum in simulations and measurements, respectively, achieving the enhancement of the conventional numerical performance by a factor of 400. Moreover, we presented additional structures to arbitrarily increase or decrease the magnetic impedance by combining different types of circuit configurations and selecting suitable circuit values. Thus, our study contributes to achieving advanced control of electromagnetic phenomena and developing applied devices by fully utilizing passive time-varying metasurfaces. 

\section*{Supplementary Material}

Supplementary Material is available for this manuscript. 

\begin{acknowledgments}
This work was supported in part by the Japan Science and Technology Agency (JST) under Fusion Oriented Research for Disruptive Science and Technology (FOREST) No.\ JPMJFR222T and the National Institute of Information and Communications Technology (NICT), Japan under the commissioned research No.\ 06201.
\end{acknowledgments}

\section*{Data Availability Statement}

The data that support the findings of this study are available from the corresponding author upon reasonable request.


\providecommand{\noopsort}[1]{}\providecommand{\singleletter}[1]{#1}%

\clearpage
\appendix
\setcounter{figure}{0}
\setcounter{table}{0}
\setcounter{equation}{0}
\setcounter{page}{1}
\renewcommand{\thetable}{S\arabic{table}}
\renewcommand{\thefigure}{S\arabic{figure}}
\renewcommand{\theequation}{S\arabic{equation}}
\renewcommand{\thepage}{S\arabic{page}}
{\Large \hspace{-4mm}\textit{Supplementary Material for}}\vspace{3mm}\\
\textbf{Passive Time-Varying Waveform-Selective Metasurfaces for Attainment of Magnetic Property Control}\vspace{3mm}\\
Yuki Kunitomo, Kairi Takimoto, Stefano Vellucci, Alessio Monti, Mirko Barbuto, Alessandro Toscano, Filiberto Bilotti, and Hiroki Wakatsuchi\vspace{3mm}\\
Email: wakatsuchi.hiroki@nitech.ac.jp

\begin{figure}[htb!]
\includegraphics[width=\linewidth]{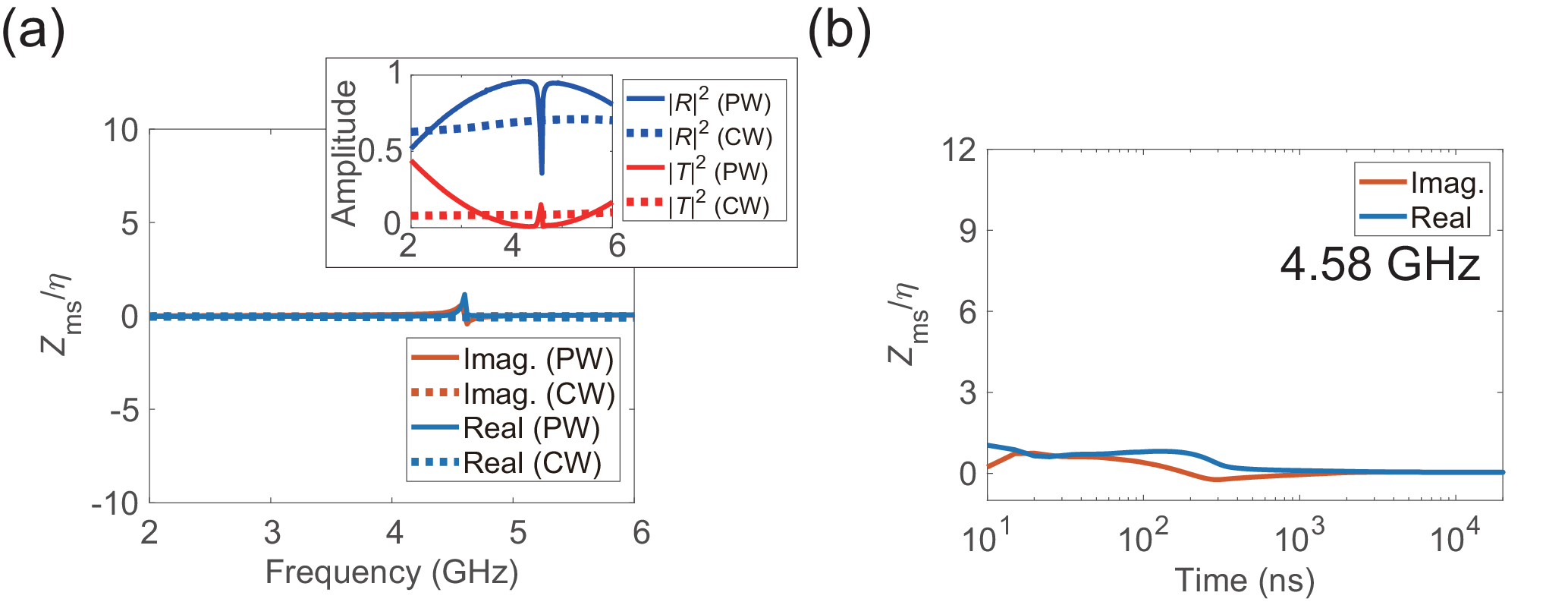}
\caption{\label{fig:figS0a}Simulated $Z_{ms}$ of the proposed two-layer structure using the original dimensions (i.e., $l=$ 5.00 mm, $h=$ 0.64 mm, and $w=$ 3.50 mm). (a) The frequency-domain result. (b) The time-domain result at 4.58 GHz. The simulation results of the improved structure are shown in Fig.\ \ref{fig1}d and Fig.\ \ref{fig1}d.}
\end{figure}

\begin{figure*}[htb!]
\includegraphics[width=\linewidth]{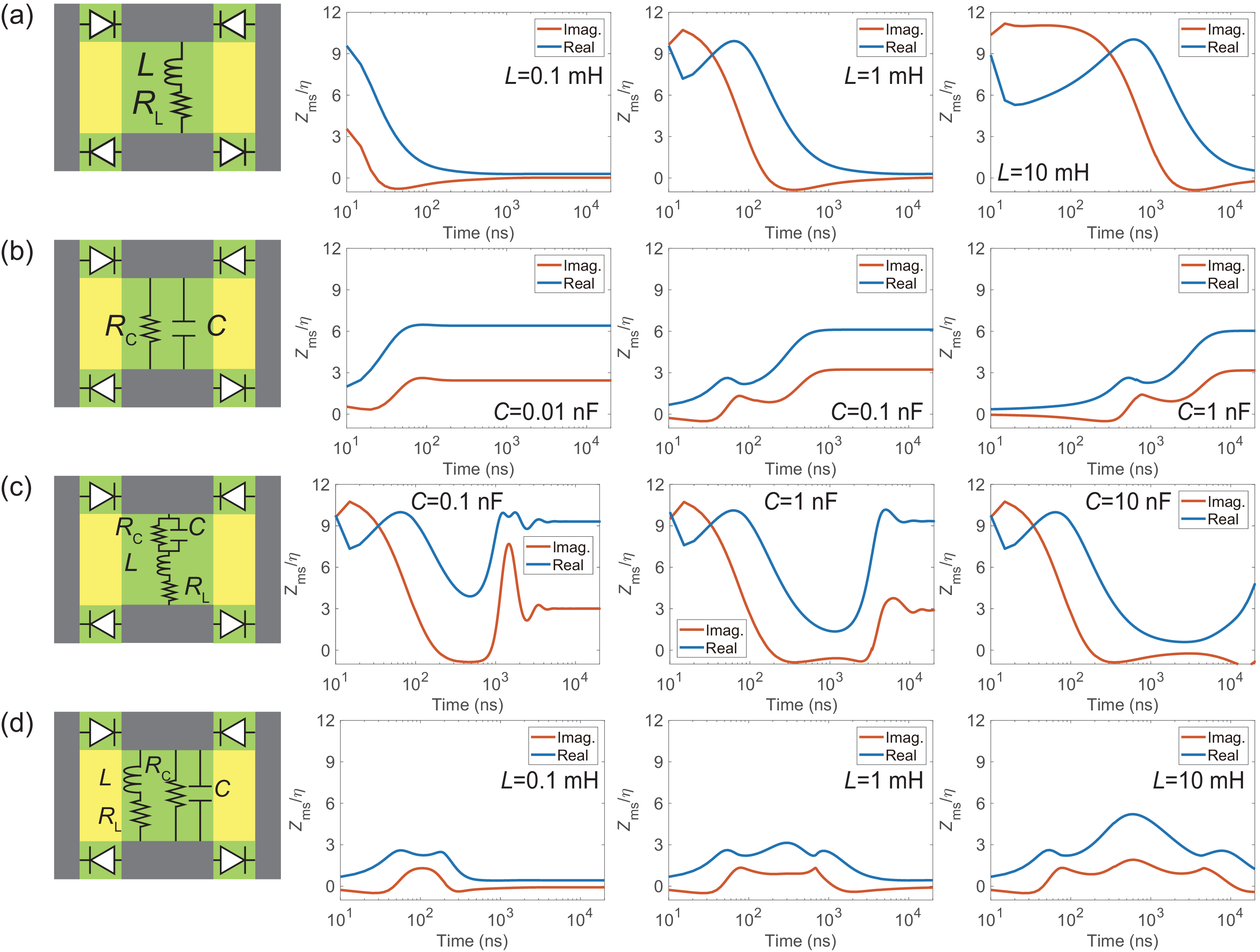}
\caption{\label{fig:figS1a}Three types of the proposed models with various circuit values. (a-d) (a) Inductor-based, (b) capacitor-based, (c) series-type, and (d) parallel-type waveform-selective metasurfaces and their $Z_{ms}$ in the time domain at 3.34 GHz. }
\end{figure*}

\begin{figure}[htb!]
\includegraphics[width=\linewidth]{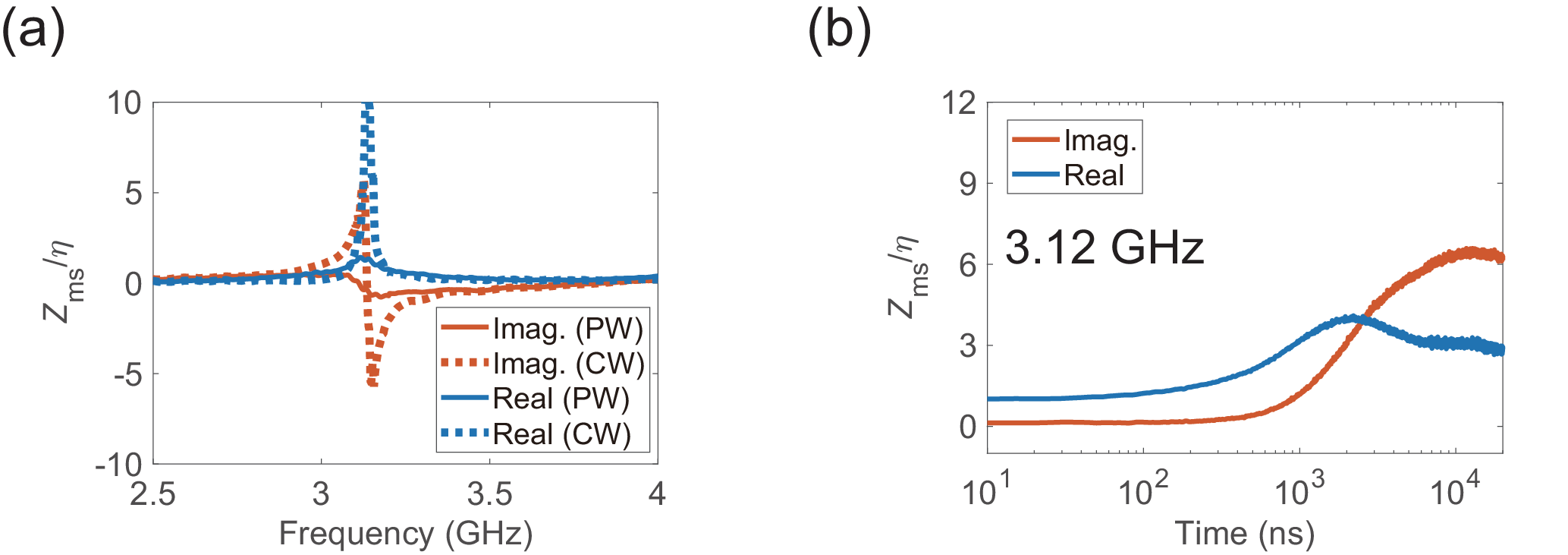}
\caption{\label{fig:figS2a}Additional experimental validation using capacitor-based waveform-selective metasurface (i.e., the top of Fig.\ \ref{fig4}a). (a) $Z_{ms}$ measured in (left) the frequency domain and (right) the time domain at 3.12 GHz. The input power was set to 13 and 15 dBm in the frequency- and the time-domain measurements, respectively, as explained in Fig.\ \ref{fig3}.}
\end{figure}

\end{document}